\documentclass{article}

\usepackage[dblblindworkshop,final]{neurips_2025}

\workshoptitle{AI for Music}

\usepackage[utf8]{inputenc} %
\usepackage[T1]{fontenc}    %
\usepackage[hidelinks]{hyperref}       %
\usepackage{url}            %
\usepackage{booktabs}       %
\usepackage{amsfonts}       %
\usepackage{nicefrac}       %
\usepackage{microtype}      %
\usepackage{xcolor}         %

\usepackage{graphicx}       %
\usepackage{natbib}         %
\usepackage{bm}             %
\usepackage{subcaption}     %
\usepackage[capitalise,noabbrev]{cleveref}

\usepackage{pifont}

\title{MusPyExpress: Extending MusPy with Enhanced Expression Text Support}

\author{
\begin{tabular}{c c c c}
    Phillip Long$^{1}$ & Hao-Wen Dong$^{2}$ & Julian McAuley$^{1}$ & Zachary Novack$^{1}$ \\
    \texttt{p1long@ucsd.edu} & \texttt{hwdong@umich.edu} & \texttt{jmcauley@ucsd.edu} & \texttt{znovack@ucsd.edu} \\
\end{tabular}
\\[1.5em]
$^{1}$Computer Science \& Engineering Department, University of California, San Diego \\
$^{2}$Department of Performing Arts Technology, University of Michigan, Ann Arbor \\[1em]
}

\begin{document}

\maketitle

\vspace{-2em}
\begin{abstract}
    Current work in modeling symbolic music primarily relies on representations extracted from MIDI-like data. While such formats allow for modeling symbolic music as sequences of notes, they omit the large space of symbolic annotations common in western sheet music broadly known as \emph{expression text}, such as tempo or dynamics, which specify time- and velocity-dependent controls on the musical composition and performance. To alleviate this gap, we present \textbf{MusPyExpress}, an extension to the popular symbolic music processing library MusPy \cite{dong2020muspy} that enables the extraction of expression text along with symbolic music for downstream modeling. Utilizing this extension, we parse the PDMX dataset \cite{long2024pdmx, xu2024generating} to illustrate the wealth of expression text available in MusicXML datasets. Additionally, we introduce multiple generative tasks, including joint expression-note generation, expression-conditioned music generation, and expression tagging, that take advantage of this additional notational information.
\end{abstract}

\vspace{-1.5em}
\section{Introduction}
\label{sec:introduction}
\vspace{-0.5em}

Over the past decade, progress in modeling symbolic music, including tasks like symbolic music generation \cite{dong2018musegan, wu2020jazz, dong2023mmt}, transcription \cite{gardner2021mt3}, and understanding \cite{jiang2019large, chou2021midibert}, has been predominantly driven by the popular MIDI \cite{rothstein1992midi} file representation. 
While MIDI is useful for modeling symbolic music as sequences of musical notes, there are a number of limitations to the encoding itself as well as the symbolic music processing libraries used to read it. Importantly, although MIDI is able to encode expression-adjacent features like time signature and tempo (in BPM), it omits most \textbf{expression text} present in western sheet music, such as tempo text (e.g., \emph{adagio}), dynamic crescendos, section markings, and note articulations.
Expression text is a key part of how music is communicated through sheet music, acting as a roadmap to the larger piece on \emph{how} the sequence of notes should be played.
In contrast to MIDI, the MusicXML format is designed with a focus on rendering sheet music for \emph{reading} rather than listening, so it often contains an abundance of this expression text.
Despite a recent surge in the number of MusicXML datasets \cite{foscarin2020asap, peter2023asap, long2024pdmx, xu2024generating}, MIDI remains the dominant standard for symbolic music data \cite{hawthorne2018enabling, raffel2016learning, wang2020pop909, donahue2019lakhnes, crestel2018database, zeng2021musicbert, ens2021building}. As a result, existing symbolic music processing libraries \cite{dong2020muspy, cuthbert2010music21, symusic2024, mckay2006jsymbolic} are generally optimized for the MIDI format and unequipped to harness the wealth of expression text available in MusicXML.

\begin{figure}
\begin{subfigure}{.5\textwidth}
  \centering
  \includegraphics[width=.95\linewidth]{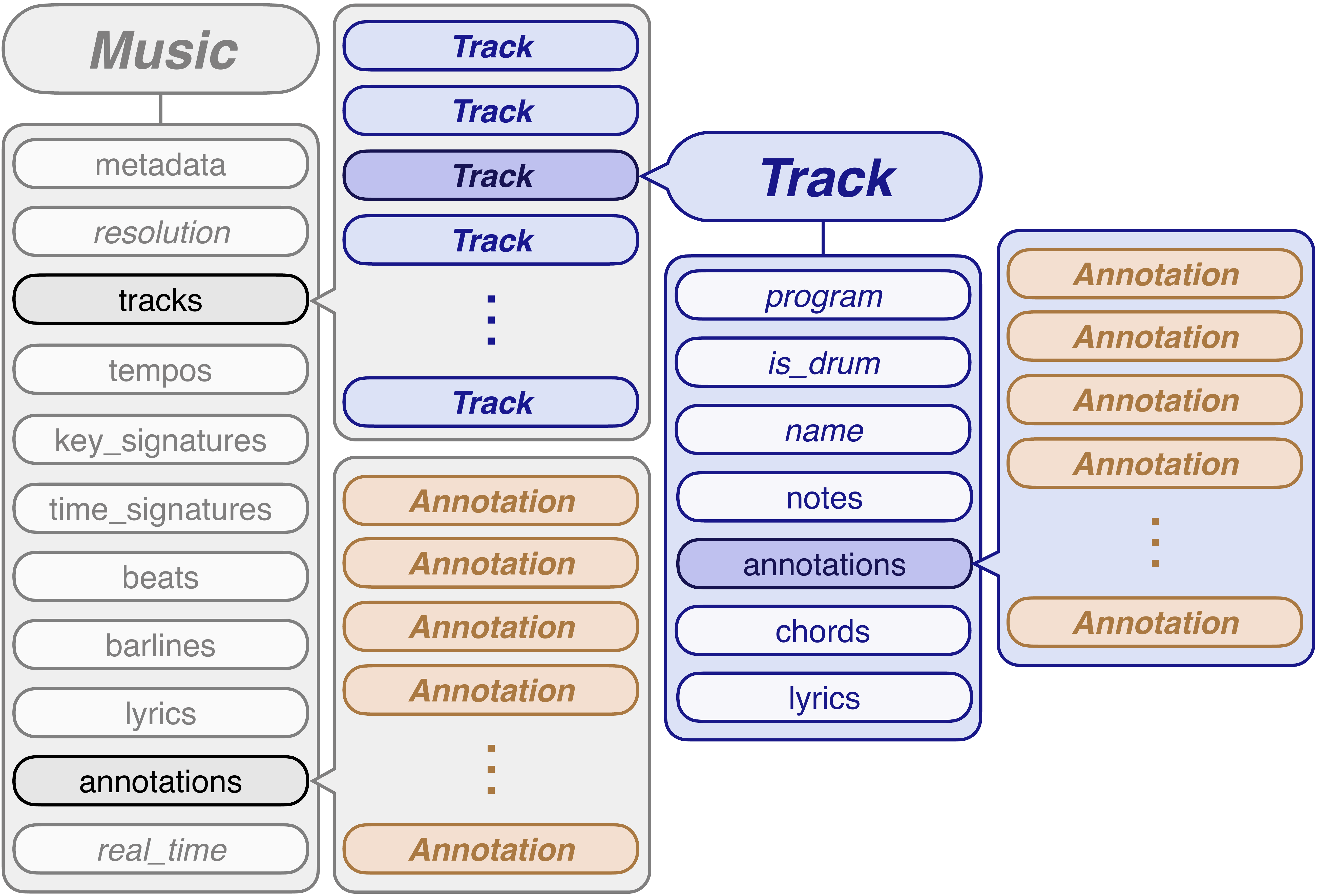}
  \caption{\emph{Music} Object Overview}
  \label{fig:MusicExpressOverview}
\end{subfigure}%
\begin{subfigure}{.5\textwidth}
  \centering
  \includegraphics[width=.95\linewidth]{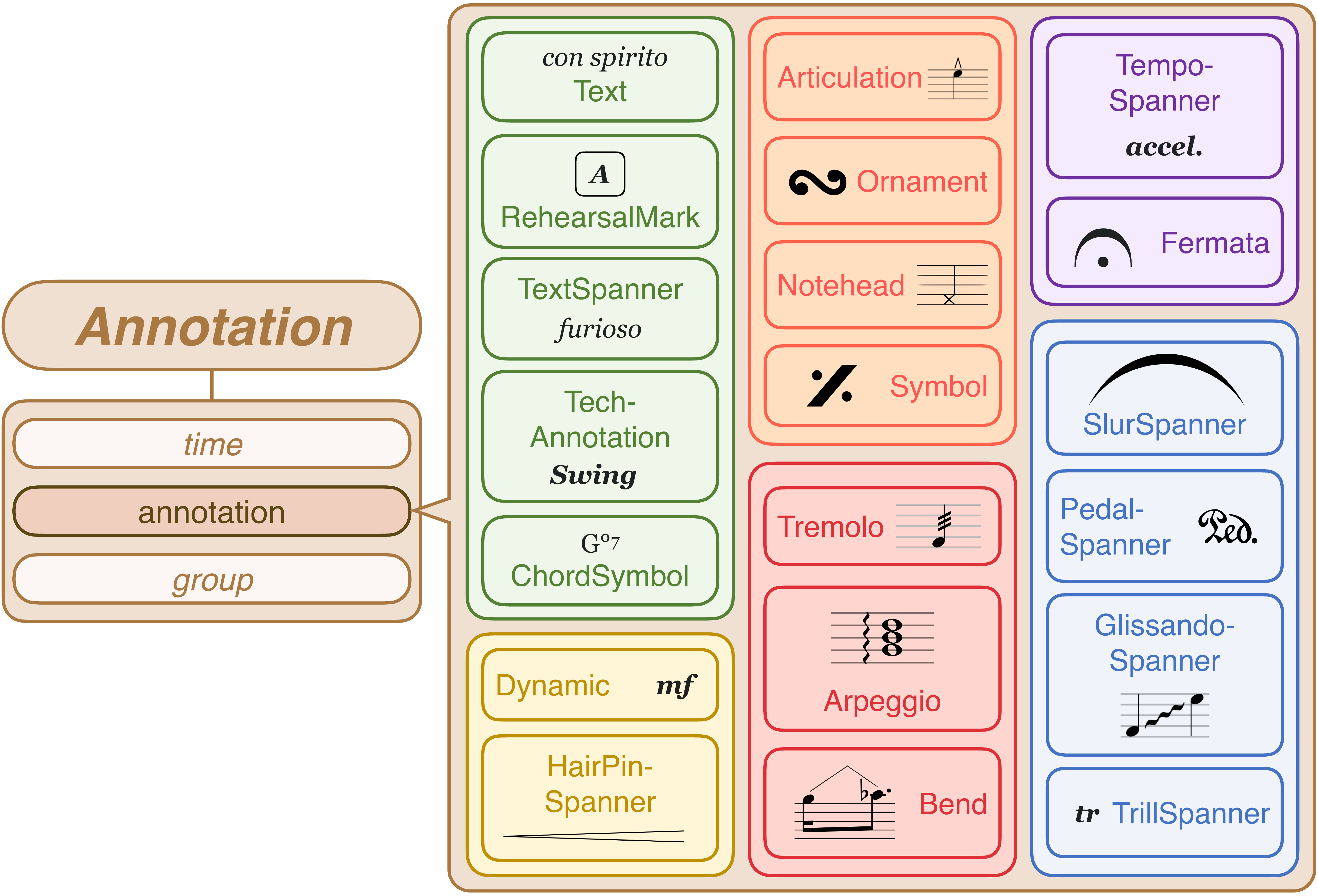}
  \caption{Annotations in MusPyExpress}
  \label{fig:MusicExpressAnnotation}
\end{subfigure}
\caption{\textbf{Overview of MusPyExpress}. Although the format of the base \emph{Music} object is the same as MusPy's \cite{dong2020muspy}, we have significantly expanded upon the annotations feature (which exists at both the score- and track-level), adding 28 new annotation types, the most of common of which are shown here. The boolean \texttt{real\_time} attribute dictates whether temporal features such as onset or duration are stored in units of time steps (metrical time) or seconds (real time).}
\label{fig:MusicExpress}
\end{figure}

To remedy this gap, we introduce \textbf{MusPyExpress}, an extension of the popular symbolic music processing library MusPy \cite{dong2020muspy} that processes expression text into discrete tensor data for downstream expression-aware sequence modeling tasks \cite{vaswani2017attention}.
As a test bed, we utilize MusPyExpress to analyze the Public Domain MusicXML (PDMX) dataset \cite{long2024pdmx, xu2024generating}, the largest available MusicXML dataset. Additionally, we show how MusPyExpress enables a number of expression-aware symbolic music tasks, including joint expression-note generation, expression-conditioned music generation, and note-to-expression sequence tagging.
MusPyExpress is available on GitHub as a branch of MusPy \footnote{\texttt{\url{https://github.com/salu133445/muspy/tree/expressive}}}.

\vspace{-1.8em}
\section{Related Works}
\label{sec:related_works}
\vspace{-1em}

Several libraries support MusicXML for symbolic music processing, but they differ in focus and capabilities. music21 \cite{cuthbert2010music21} is one of the most well-known toolkits, and despite extensive support for parsing expression text, these annotations are often embedded deep within its object hierarchy, making them cumbersome to extract for downstream modeling tasks. symusic \cite{symusic2024} also supports MusicXML, but its primary design goal is efficient MIDI parsing rather than broad symbolic music representation, and the library was not developed with an emphasis on capturing expression text. jSymbolic \cite{mckay2006jsymbolic} focuses on extracting statistical features from symbolic music, making it valuable for classification tasks but less suited for flexible input-output handling across formats. Despite these viable alternatives, our MusPyExpress library not only supports MusicXML but also emphasizes exposing expression text in a Python-based, lightweight, and accessible manner to facilitate downstream modeling.

\vspace{-0.4em}
\section{MusPyExpress}
\label{sec:musicexpress}
\vspace{-1em}

We design an extension of the Python library MusPy \cite{dong2020muspy}, a package built to handle various forms of symbolic music with a universal \emph{Music} object, which we term \textbf{MusPyExpress}. Given a piece of symbolic music, the resulting Music object can contain information on song layout, metadata, individual parts, time signatures, key signatures, tempo markings, notes, and bar boundaries. However, MusPy fails to parse most expression text, like dynamics, slurs, and articulations, and we design MusPyExpress to address this shortcoming.
A Music object has the capacity to store ``annotations,'' a catchall term for any information unrelated to the notes in a song or track. MusPyExpress expands this functionality, and creates a multitude of different Python objects, representing different types of expression text summarized in \cref{fig:MusicExpress},
meant for storage within an annotation. Annotations can be stored in two places: at the score-wide level, generally reserved for temporally-related expression text and rehearsal marks, or the track-specific level, where dynamics and individual text instructions reside. In our implementation, we use the term ``spanner'' to refer to expression text that ``spans'' an explicit duration. For instance, a \emph{TempoSpanner} object, like an \emph{accelerando}, differs from a \emph{Tempo} object, such as a \emph{vivace} marking. Because MusicXML does not constrain creators to a strict set of expression text options, we have implemented the collection of the most common types of expression text. We further discuss the contributions of MusPyExpress in \cref{app:contributions_of_musicexpress}.

\vspace{-0.4em}
\section{Dataset Analysis}
\label{sec:dataset_analysis}
\vspace{-1em}

\begin{figure}
\begin{subfigure}{.5\textwidth}
  \centering
  \includegraphics[width=.95\linewidth]{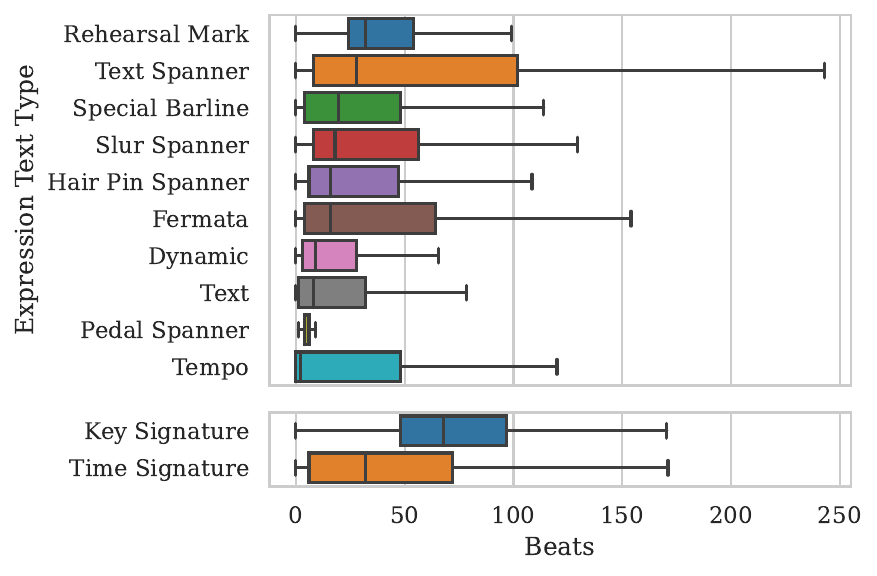}
  \caption{Relative Density of Expression Text}
  \label{fig:dataset_analysis_relative_density}
\end{subfigure}%
\begin{subfigure}{.5\textwidth}
  \centering
  \includegraphics[width=.95\linewidth]{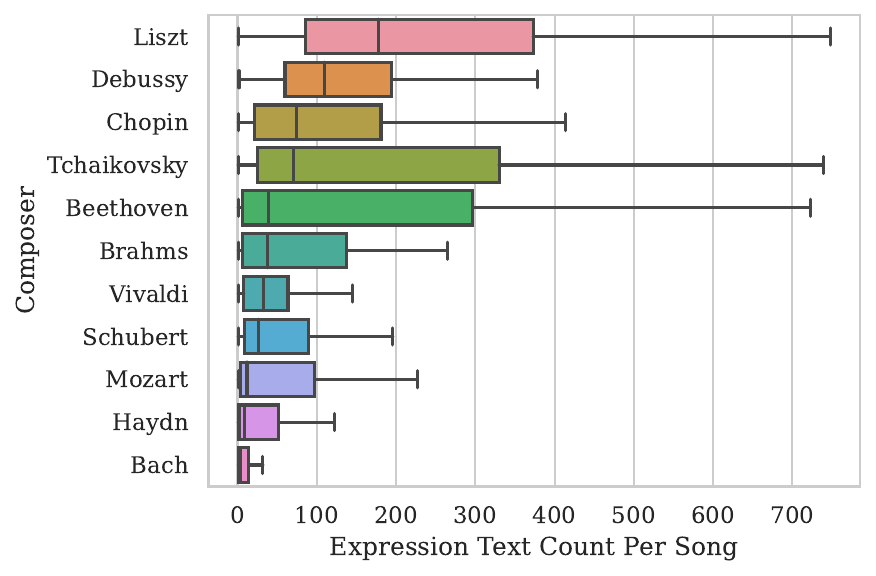}
  \caption{Expression Text Usage by Composer}
  \label{fig:dataset_analysis_composers}
\end{subfigure}
\caption{\textbf{Expression Text in PDMX}. \Cref{fig:dataset_analysis_relative_density} shows boxplots of relative expression text density. 
Throughout this work, ``Special Barline'' refers to special barline types like double and dotted barlines and excludes the ordinary single barline. We include the relative density of key and time signature changes for reference. \Cref{fig:dataset_analysis_composers} shows boxplots of expression text usage (i.e.~the number of different expression text markings) per song for a select group of composers.}
\label{fig:dataset_analysis}
\end{figure}

We utilize the Public Domain MusicXML (PDMX) dataset \cite{long2024pdmx, xu2024generating} as a test bed for MusPyExpress in order to illustrate the wealth of available expression text in MusicXML
datasets. 
Of the original 222,820 MusicXML files in PDMX, 212,406 (95.33\%) contain expression text, comprising $>$3.5M total expression text markings and averaging 20.1 markings per song. We include further analysis of PDMX's expression text distribution in \cref{app:musicexpress_on_pdmx}.

The average \emph{absolute} expression text density (i.e.~the metrical distance between two expression text markings, regardless of type) in PDMX is 17.21 beats. Meanwhile, the \emph{relative} density over time for each expression text type (i.e.~the distance in beats between two expression markings of the same type) is described in \cref{fig:dataset_analysis_relative_density}. Songs tend to avoid overusing structural control expression types, as doing so diminishes their effect, while expression types like tempos, dynamics, and text features can be quite concentrated, even more so for note-level annotations like pedal markings. Rehearsal marks exhibit a sparser distribution than their normal text counterparts, as these markings are reserved for demarcating different sections of a song.
We also make use of PDMX's composer metadata to analyze expression text usage by composer in \cref{fig:dataset_analysis_composers}, finding that Romantic composers include expression text more frequently than their Baroque and Classical counterparts. This aligns with historical practice, since Romantic composers often explicitly encoded expression markings to emphasize individual expressivity, while Baroque and Classical composers often relied more on performance conventions of their time, resulting in fewer expression markings.

\vspace{-0.4em}
\section{Experiments}
\label{experiments}
\vspace{-1em}

\begin{table}[h!]
\begin{center}
\footnotesize\begin{tabular}{l|ccc|c}
    \toprule
     Model & $\textbf{PCE}$ & $\textbf{SC}$ (\%) & $\textbf{GC}$ (\%) & $\textbf{Perplexity}$ ($\downarrow$) \\
    \midrule
    Ground Truth & $2.68 \pm 0.00$ & $97.40 \pm 0.01$ & $93.67 \pm 0.01$ & - \\
    \midrule
    Baseline (\emph{M}) & $2.82 \pm 0.02$ & $93.31 \pm 0.45$ & $\mathbf{93.87 \pm 0.21}$ & $2.80$ \\
    Joint, Prefix (\emph{M}) & $1.75 \pm 0.07$ & $95.57 \pm 0.45$ & $97.41 \pm 0.19$ & $\textbf{2.64}$ \\
    Joint, Anticipation (\emph{M}) & $\mathbf{2.72 \pm 0.03}$ & $92.77 \pm 0.46$ & $93.93 \pm 0.26$ & $3.58$ \\
    Conditional, Prefix (\emph{M}) & $1.83 \pm 0.06$ & $94.91 \pm 0.44$ & $94.43 \pm 0.32$ & $3.07$\\
    Conditional, Anticipation (\emph{M}) & $1.58 \pm 0.05$ & $\mathbf{96.72 \pm 0.34}$ & $97.80 \pm 0.14$ & $3.02$\\
    \midrule
    Baseline (\emph{R}) & $\mathbf{2.56 \pm 0.03}$ & $95.87 \pm 0.36$ & $92.19 \pm 0.37$ & $4.38$ \\
    Joint, Prefix (\emph{R}) & $1.74 \pm 0.07$ & $\mathbf{98.61 \pm 0.17}$ & $95.50 \pm 0.36$ & $3.79$ \\
    Joint, Anticipation (\emph{R}) & $2.22 \pm 0.05$ & $95.16 \pm 0.49$ & $95.47 \pm 0.28$ & $\textbf{3.52}$ \\
    Conditional, Prefix (\emph{R}) & $2.37 \pm 0.05$ & $92.93 \pm 0.44$ & $92.17 \pm 0.48$ & $4.05$\\
    Conditional, Anticipation (\emph{R}) & $2.40 \pm 0.05$ & $94.08 \pm 0.36$ & $\mathbf{94.48 \pm 0.31}$ & $4.43$ \\
    \bottomrule
\end{tabular}
\end{center}
\caption{\textbf{Evaluation Results}. Objective evaluation results for joint (1) and conditional (2) generation tasks. We evaluate the same metrics proposed in Multitrack Music Transformer (MMT) \cite{dong2023mmt} -- pitch class entropy (PCE), scale consistency (SC) and groove consistency (GC) -- where a closer value to that of the ground truth is considered better. We also evaluate each model's perplexity over notes (i.e.~ignoring expression text), where lower is better.
}
\label{tab:results}
\end{table}

MusPyExpress enables a number of symbolic music modeling tasks that take into account the sequence of expression text features. We focus on modeling specific \emph{tracks} rather than full scores, which we leave for future work. At a high level, a track is comprised of a sequence of notes $\bm{n} = (\bm{n}_{1:N})$ and expression text tokens $\bm{e} = (\bm{e}_{1:M})$. Using the subset of PDMX \emph{tracks} that contain annotations,
we propose three different generation tasks within this framework:
\begin{enumerate}
    \item \textbf{Joint Note-Expression Text Generation}: We model \emph{both} sequences together $p_\theta(\bm{n}, \bm{e})$ and thus produce a model that can generate both notes and expression text. This task mirrors how composers actually write music, adding new notes and expression text simultaneously.
    
    \item \textbf{Expression-Conditioned Note Generation}: We modify the standard symbolic music generation task with expression text annotations $p_\theta(\bm{n} \mid \bm{e})$, learning a model to produce notes \emph{given} a sequence of expression text tokens. This task could prove useful in film music generation and sound design for games and audiobooks, where it is often useful to match music to the climax, tension, and rhythm of the inputs.
    
    \item \textbf{Expression Tagging}: Given an existing note sequence $\bm{n}$, we learn a model $p_\theta(\bm{e} \mid \bm{n})$ that tags the sequence with expression text $\bm{e}$ (i.e. add expression text to a planned musical score). This task serves as an intermediate step in the MIDI to performance rendering of a musical score; given only a plain MIDI file, a trained model could suggest appropriate expression markings such as \emph{legato}, \emph{crescendo}, or \emph{dolce}, enabling automatic annotation of extensive collections of unexpressive MIDI data widely available in large-scale MIDI datasets.
\end{enumerate}
We refer to the second and third experiments as ``conditional,'' and each experiment is performed in both the metrical (\emph{M}) and real (\emph{R}) time domains. We discuss different data representations used, interleaving strategies, and experimental setup for each task in \cref{app:experiment_details}.

\vspace{-0.8em}
\paragraph{Joint/Expression-Conditioned Results}
\label{joint_conditioned_results}

\Cref{tab:results} compares our models trained on different timing and conditioning schemes against a \emph{note-only} MMT \cite{dong2023mmt} baseline. No model configuration clearly dominates across all metrics (which may be an artifact of our small model size).
However, we find our best overall perplexity (i.e.~a model's certainty in its predictions) in the joint prefix-conditioned metrical-time model, providing evidence that expression tokens can provide useful local information for realistic music generation.

\vspace{-0.8em}
\paragraph{Expression Tagging Results}
\label{expression_tagging_results}

\begin{table}
    \centering\footnotesize
    \begin{tabular}{l|ccccc|c}
    \toprule
    Model & \textbf{Beat} & \textbf{Position} & \textbf{Time} & \textbf{Value} & \textbf{Duration} & \textbf{Total} \\
    \midrule
    Baseline ($\it{M}$) & 11.36 & 86.34 & - & 1.26 & 14.68 & 0.05 \\
    Prefix ($\it{M}$) & \textbf{62.80} & \textbf{93.55} & - & \textbf{55.60} & \textbf{54.15} & \textbf{29.41} \\
    Anticipation ($\it{M}$) & 59.78 & 92.73 & - & 50.43 & 50.11 & 24.41 \\
    \midrule
    Baseline ($\it{R}$) & - & - & 4.17 & 13.86 & 59.02 & 1.63 \\
    Prefix ($\it{R}$) & - & - & \textbf{46.05} & \textbf{57.39} & \textbf{66.54} & \textbf{33.90} \\
    Anticipation ($\it{R}$) & - & - & 40.25 & 53.32 & 65.60 & 27.75 \\
    \bottomrule
    \end{tabular}
    \vspace{1em}
    \caption{\textbf{Expression Tagging Results}. Per-field accuracy of expression tagging (3) models on validation set. As a baseline, we examine a model that simply predicts the most common value of each field. 
    Baseline differences between the metrical and real timing schemes may arise because the real time representation truncates songs at 60 seconds, excluding expression text in longer pieces and thereby altering the distribution of field values.
    }
    \label{tab:econditional}
\end{table}

In \cref{tab:econditional}, we show accuracy results for expression tagging with both timing and interleaving schemes. Notably, we find that the prefix-conditioned models, especially those in real time, clearly dominate the anticipatory models, as the prefix setup is much simpler when predicting the sparse sequence of expression tokens (which anticipatory models can struggle on \cite{thickstun2023anticipatory}).

\vspace{-0.8em}
\section{Conclusion}
\label{sec:conclusion}
\vspace{-1em}

We present \textbf{MusPyExpress}, an extension to the popular symbolic music processing library MusPy \cite{dong2020muspy} that supports expression text processing for use in downstream modeling tasks. We use the PDMX \cite{long2024pdmx, xu2024generating} dataset as a test bed for our extension to show the abundance of available expression text in MusicXML 
datasets. We then employ the subset of PDMX tracks containing these annotations for three generative tasks -- joint note-expression text generation, expression-conditioned note generation, and expression tagging -- illustrating how MusPyExpress enables expression-aware downstream modeling. In the future, we plan to further explore expression-tagging systems that would allow us to annotate large, unexpressive MIDI datasets with expression text recommendations, as well as expand expression-conditioned generation into full fine-grained text-to-music generation.

\bibliographystyle{plain}
\bibliography{bibliography}

\newpage
\appendix

\section{Contributions of MusPyExpress}
\label{app:contributions_of_musicexpress}

\subsection{Type of Expression Text}
\label{app:expression_text_types}

MusPyExpress introduces 28 Python classes, representing different expression text types commonly found in western symbolic music. These classes are meant for storage as the \texttt{annotation} attribute inside of MusPy's more general \emph{Annotation} object. Broadly, we can group these expression text types into four categories: text, structural controls, note-level annotations, and general-purpose abstractions.  

\textbf{Text-based expressions} include rehearsal marks and natural language annotations such as \emph{con spirito}.  
\begin{itemize}
    \item \emph{ChordSymbol}: chord labels such as ``Cmaj7.''  
    \item \emph{RehearsalMark}: rehearsal indicators like ``A'' or ``Verse.''
    \item \emph{Text}: general text annotations.  
    \item \emph{TextSpanner}: extended text annotations spanning multiple notes.  
\end{itemize}

\textbf{Structural controls} encompass the broadest range, including tempo and dynamic markings, time and key signature changes, barlines, and \emph{fermatas}.  
\begin{itemize}
    \item \emph{ChordLine}: horizontal line indicating repeated chord structure.  
    \item \emph{Dynamic}: dynamic markings such as \emph{mf} or \emph{ff}.  
    \item \emph{Fermata}: sustained pause on a note or rest.  
    \item \emph{HairPinSpanner}: crescendo or diminuendo markings.  
    \item \emph{OttavaSpanner}: notation for playing passages one or more octaves higher or lower.  
    \item \emph{Symbol}: general score symbol or other graphical markers.  
    \item \emph{TempoSpanner}: tempo markings spanning multiple notes such as \emph{ritardando} and \emph{accelerando}.  
\end{itemize}

\textbf{Note-level annotations} include articulations, slurs, and pedal markings, which can indicate unique stylistic passages when found in clusters.  
\begin{itemize}
    \item \emph{Arpeggio}: broken chord indication.  
    \item \emph{Articulation}: markings such as \emph{staccato} or accent.  
    \item \emph{Bend}: pitch bend for fretted instruments.  
    \item \emph{GlissandoSpanner}: sliding motion between two notes.  
    \item \emph{Notehead}: alternate notehead shapes.  
    \item \emph{Ornament}: decorative figures like mordents or turns.  
    \item \emph{PedalSpanner}: sustain pedal or similar keyboard pedal markings.  
    \item \emph{SlurSpanner}: slurs connecting multiple notes.  
    \item \emph{TechAnnotation}: instrument-specific technique annotations such as \emph{mute} for a trumpet.  
    \item \emph{Tremolo}: rapid repetition of a note or alternation between notes.  
    \item \emph{TremoloBar}: whammy-bar notations for guitar.  
    \item \emph{TrillSpanner}: rapid alternation between two adjacent notes.  
    \item \emph{VibratoSpanner}: vibrato or wavy-line indications.  
\end{itemize}

\textbf{General-purpose abstractions} are more for internal organization and storing miscellaneous expression text that do not fall neatly into the above categories.  
\begin{itemize}
    \item \emph{Point}: structural indicator for positioning in a score.  
    \item \emph{Spanner}: general-purpose spanning annotation.  
    \item \emph{Subtype}: subtype label for higher-level categories.  
    \item \emph{SubtypeSpanner}: subtype annotation spanning multiple measures.  
\end{itemize}

Beyond annotations, MusPyExpress also extends several of MusPy's base classes to better capture score-level details. In particular, it adds support for tempo marking text (in addition to raw BPM values), barline types for tracking special double or dotted barlines, and a grace-note attribute for notes to indicate \emph{appoggiaturas} and \emph{acciaccaturas}.

\subsection{Timing}
\label{app:timing}

In MusPy, time is stored with a metrical time unit known as time steps. Each time step represents a fraction of a beat, and is therefore tempo agnostic. However, use often necessitates seconds-based encoding, so MusPyExpress includes functionality to convert to and from metrical and real time in a way that accounts for any temporally-related expression text in the music. This is valuable not only for realizing a score into MIDI or audio, but also for downstream modeling tasks where real time encoding makes it easier to align and interpret expression markings that are difficult to capture in tempo-agnostic time steps, such as tempo changes and \emph{fermatas}. By supporting both metrical and real time representations, MusPyExpress provides more flexible handling of expression-aware symbolic music modeling.

\subsection{File I/O}
\label{app:file_io}

MusPyExpress also improves on MusPy’s audio- and symbolic-output capabilities. MusPy was built to process a variety of symbolic music types, and has the ability to render these various file types as MIDI (symbolic domain) and WAV (audio domain). Native MusPy renderings are limited to notes with velocities inherited from the symbolic music source and, importantly, do not reflect any expression text dynamic markings. While MusPy renderings realize global tempo changes (e.g., BPM specifications), they do not account for local tempo variations such as \emph{ritardando} or \emph{accelerando}.

In MusPyExpress, audio and symbolic domain outputs now reflect any expression text markings present in a given Music object. For instance, note durations for a slurred musical section extend to the next note, and accented notes are markedly louder than their non-accented peers. Tempo-related expression text, like \emph{ritardandos}, and volume-related features, like \emph{crescendos}, are realized through changes in tempo and velocity, respectively. To translate such markings into concrete values, we use a set of configurable constants (e.g., slowing tempo 3$\times$ for fermatas or increasing velocity 1.5$\times$ for accents), which default to our manually curated values, in order to consistently map expression text into numerical adjustments.

Similar to MusPy, a MusPyExpress object can both import-from and export-to JSON format, which allows storage of these objects without any information loss.

\section{Analyzing PDMX with MusPyExpress}
\label{app:musicexpress_on_pdmx}

\subsection{Parsing MusicXML-Adjacent Formats}
\label{app:musicxml_adjacent}

MusPyExpress includes support for parsing not only standard MusicXML files but also MusicXML-adjacent formats that share similar structure and semantics. One such format is MSCZ (used by MuseScore), which differs only slightly in terminology from MusicXML but is otherwise structurally compatible. To demonstrate this, in addition to PDMX's MusicXML files, we also parse the original MSCZ files from which PDMX was derived to illustrate that MusPyExpress can read other symbolic music formats containing expression text.

\begin{table}
    \centering
    \begin{tabular}{r|c}
    \toprule
    \textbf{Expression Text Type} & \textbf{Count} \\
    \midrule
    Tempo & 1,082,640 \\
    Dynamic & 1,069,492 \\
    Text & 467,479 \\
    Barline & 336,029 \\
    Rehearsal Mark & 176,579 \\
    Hair Pin Spanner & 164,014 \\
    Slur Spanner & 86,764 \\
    Fermata & 78,572 \\
    Pedal Spanner & 48,421 \\
    Text Spanner & 3,651 \\
    \midrule
    \textbf{Total} & 3,513,641 \\
    \bottomrule
    \end{tabular}
    \vspace{1em}
    \caption{\textbf{Expression Text Frequencies}. Frequencies of different expression text types in PDMX. Tempo and dynamic markings are the most common, as every song requires at least one of each, while pedal and text spanners are the least frequent.}
    \label{tab:dataset_analysis_counts}
\end{table}

\begin{figure}
\centerline{
\includegraphics[width=0.5\textwidth]{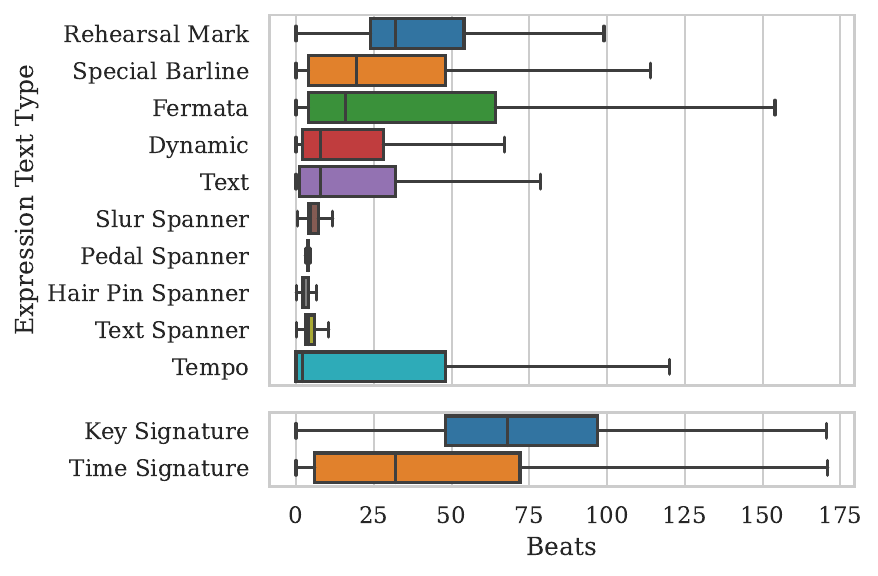}}
\caption{\textbf{Expression Text Durations}. Boxplots of durations (in beats) for different categories of expression text. Key signatures, rehearsal marks, and time signatures tend to span entire sections of songs, and thus have some of the longest durations. Pedal and text spanners, meanwhile, usually only apply to short sequences of notes and so are often quite brief. Tempo markings exhibit a long tail of extended durations, likely because many songs contain frequent tempo changes resulting in short local spans, while a few songs may have an initial tempo that persists across the entire piece.}
\label{fig:dataset_analysis_durations}
\end{figure}

\subsection{Expression Text Usage in PDMX}
\label{app:pdmx_expression_text_usage}

PDMX contains $>$3.5M total expression text annotations. In \cref{tab:dataset_analysis_counts}, we dissect this total into counts for each available expression text type. Tempo and dynamic markings are the most prevalent categories, since every song requires at least one of each, whereas pedal and text spanners occur comparatively infrequently.

While some expression text types in PDMX have an explicit duration attribute, many, such as time and key signatures, do not, and so to analyze their distribution, we examine their implied duration, defined as the metrical distance to the next expression text of the same type, or to the end of the song if none follow. \Cref{fig:dataset_analysis_durations} shows boxplots of of these explicit and implied durations (in beats) for different expression text types. Expression types with implied durations, including dynamics, \emph{fermatas}, and rehearsal marks, generally span entire sections and thus have longer durations, whereas explicitly-delimited markings like text spanners and pedal markings exhibit shorter lengths. For expression text types with implied duration, relative density (the metrical distance from one expression text to the next of the same type, examined in \cref{fig:dataset_analysis_relative_density}) is identical, and only in the case of explicit duration are relative density and implied duration different.

\section{Experiment Details}
\label{app:experiment_details}

\subsection{Data Representation}
\label{app:data_representation}

In this work, we use the tokenization scheme from Multitrack Music Transformer (MMT) \cite{dong2023mmt}. Namely, we represent a song as a sequence of events $\bm{x}_{i}$ (either a note $\bm{n}_i$ or expression $\bm{e}_i$), where each event is represented by a tuple of variables. MMT uses a metrical timing system, where $\bm{x}_{i}$ has six fields: type, beat, position, pitch, duration, instrument. Songs in MMT encoding have a structural ``preamble'' starting each sequence, encoding the list of instruments and the start of notes.

To adapt the MMT framework for expression-aware modeling, we make a number of modifications. Most saliently, we encode expression text in the MMT framework by including an additional value in the type field to denote expression tokens, and convert MMT's pitch field into a \emph{value} field, which combines the standard vocabulary of 128 MIDI-pitches with nearly 700 expression text values (i.e.~an expression token is a note token where its ``pitch'' denotes the type of expression text).
Additionally, we provide an alternative \emph{real} time encoding, where the beat and position fields are replaced by a single real time field, and temporally-related fields, namely time and duration, are encoded in seconds. We do this as many expression text types are temporal in nature, which would not be captured at the note-level in a metric system. Furthermore,  we add a velocity field to support the learning of dynamic-related expressions.

\subsection{Interleaving Methods}
\label{app:interleaving_methods}

As we seek to model the sequence of notes and expression text together with an \emph{autoregressive} model, how we interleave the notes and expression text into a single sequence $\bm{x} = \texttt{interleave}(\bm{n}, \bm{e})$ can make a large difference in performance \cite{thickstun2023anticipatory}. Under each conditional experiment, ``events'' are conditioned on ``controls.'' For example, expression text \emph{controls} dictate the note \emph{events} generated by an expression-conditioned model. 
In this work we test two different interleaving methods: prefix and anticipatory conditioning.

Prefix conditioning, used in most encoder-decoder models \cite{zhu2018xiaoice, ren2020popmag}, sets the control sequence as a \emph{prefix} to the main event sequence. While simple, this forces the model to attend to long-term dependencies in order to model the relationship between events and controls that are close in time. Anticipation \cite{thickstun2023anticipatory} seeks to remedy this fact and place controls close in the sequence to the events they affect. Specifically, under anticipatory conditioning, 
a control with onset time $t$ is placed after the first event $i$ with onset time $t_{i}$ such that $t_{i} \geq t - \delta$ for some offset $\delta$, which effectively interleaves the controls \emph{within} the event sequence rather than separate from it (see \cite{thickstun2023anticipatory} for an in depth description). 
For all experiments, we set $\delta = 8$, which is interpreted under metrical and real time schemes as beats and seconds, respectively.

\subsection{Experimental Setup and Metrics}
\label{app:experimental_setup_and_metrics}

For all experiments, we use an MMT-style decoder-only transformer, with 6 layers, 8 attention heads and a hidden dimension of 512, totaling at about 20M parameters. We use an absolute positional embedding and set the maximum sequence length to 1024. We train each model for 80K steps with a batch size of 8, learning rate of 5e-4, with all experiments performed on a single NVIDIA GeForce RTX 3090 GPU, and use the Adam optimizer with default hyperparameters. For the joint and expression-conditioned experiments, we set the expression tokens as the controls for both interleaving methods. For expression tagging, we flip the roles and set the \emph{notes} as the controls for interleaving. Note that all experiments are trained with the same next token prediction loss, the only difference being that the loss values of ``control'' tokens are zero-d out for the expression-conditioned (expression text) and expression tagging (notes) experiments. We use an 80\%-10\%-10\% train-val-test split on the subset of 357,749 PDMX tracks that include at least one annotation.

For the joint and expression-conditioned generation tasks, we follow past work \cite{dong2023mmt, wu2020jazz, thickstun2023anticipatory} and report the pitch class entropy, scale consistency, and groove consistency across 256 generations per model configuration, which measure how well the model captures the underlying musical patterns of the data, as well as the perplexity over notes (i.e.~ignoring expression text) on the held out test set. Expression sequences for the expression-conditioned generations are pulled randomly from the test set. For the tagging task, we report per-field accuracy on predicted expression text, ignoring the type, instrument, and velocity fields, which generally achieve 100\% accuracy by nature since our data representation encodes these fields as constant values in expression text events within single-track sequences.

\end{document}